\documentclass[preprint,pre,aps]{revtex4-2}
\usepackage[dvips]{graphicx}
\usepackage[usenames,dvipsnames]{color}
\bibstyle{prsty}

\begin{document}
 \title{Mesoscopic theory for a double layer capacitance in concentrated ionic systems 
 }
	\author{  A. Ciach\footnote{ the email address of the corresponding author:  aciach@ichf.edu.pl }}
\affiliation{Institute of Physical Chemistry,  Polish Academy of Sciences, 01-224 Warszawa, Poland, }
\author{ O. Patsahan}
\affiliation{Institute for Condensed Matter Physics of the National
	Academy of Sciences of Ukraine, 1 Svientsitskii St., 79011 Lviv,
	Ukraine}
\date{\today} 
\begin{abstract}
Effect of an oscillatory decay of the charge density  in concentrated ionic solutions and ionic liquids 
on the double-layer capacitance  is studied in a framework of a mesoscopic theory. Only Coulomb and steric forces between the ions  that are present in all ionic systems are taken into account. We show that the  charge oscillations lead to a rescaled distance between the electrode and the virtual 
 monolayer of counterions in the Helmholtz capacitance, and the scaling factor depends on the period of the charge oscillations.  
  Our very simple formula  for large density of ions and small voltage  can serve as a reference point for the double layer capacitance in concentrated ionic solutions and ionic liquids, and can help to disentangle the universal and specific contributions to the capacitance in particular systems.
\end{abstract} 
\maketitle
\section{introduction}
Understanding of fundamental  properties of a double-layer in concentrated electrolytes and room temperature ionic liquids is of great importance for 
designing  fuel cells, batteries, supercpacitors and energy storage devices.    
In the early model of the double layer, the counterions form a layer separated from the electrode by the distance $L$ that was first approximated by the ionic diameter $a$. In this model, the Helmholtz capacitance is simply $C_H=\epsilon/(4\pi a)$, where $\epsilon$ is the dielectric constant. In a more accurate model of the double layer~\cite{debye:23:0}, thermal motion of the ions is taken into account, and in the screening cloud of ions  the charge density decays with the distance from the electrode with the 
 Debye screening length $\lambda_D$~\cite{debye:23:0,barrat:03:0}.
 In the Debye capacitance, the diffuse layer of counterions is taken into account, and $L=\lambda_D$. 
The Debye capacitance $C_D=\epsilon/(4\pi \lambda_D)$ is a reference point for dilute electrolytes at very small voltages.

In concentrated electrolytes and ionic liquids (IL), however, the charge density decays with the distance from the electrode in an oscillatory way, and the decay length differs significantly from $\lambda_D$. The simple picture of dilute electrolyte where the average distance between the ions is much larger than their diameter is no longer valid, and the assumption of point-like ions is not justified.

A lot of effort has gone into experimental, 
theoretical and simulation 
studies of the structure of concentrated electrolytes and IL \cite{Gebbie2013,smith:16:0,lee:17:0,lee2017,Kjellander2018,groves:24:0,Coupette2018,
Rotenberg_2018,Adar2019,
Coles2020,Zeman2020,Zeman2021,Cats2021,safran:23:0,Ciach2023,Yang2023,Hrtel2023,wang:24:0,Elliott2024,ciach:21:0}. On the one hand, universal behavior was observed by the Perkin group for the decay length $\lambda_s$ of the force between charged objects immersed in concentrated electrolytes or IL~\cite{smith:16:0,lee:17:0,lee2017,groves:24:0}.   $\lambda_s$ determined in these experiments obeys the scaling $\lambda_s/\lambda_D\sim(a/\lambda_D)^n$ with $n=3$ for all  ionic systems confined between crossed mica cylinders~\cite{smith:16:0,lee:17:0,lee2017,groves:24:0}. This result indicates increasing charge-charge correlation length  for increasing concentration of ions.
In theoretical and simulation studies of concentrated ionic solutions as well as in recent SAXS experiments~\cite{Rotenberg_2018,Adar2019,
Coles2020,Zeman2020,Zeman2021,Cats2021,safran:23:0,Ciach2023,Yang2023,Hrtel2023,wang:24:0,Elliott2024,ciach:21:0,dinapajooh:24:0}, the increasing $\lambda_s$  for increasing concentration of ions was confirmed, but different values of $n$ were obtained in different works. The underscreening observed in the above experimental, simulation and theoretical studies was not confirmed by the AFM experiments with ionic system confined between silica surfaces~\cite{kumar:22:0}, and the issue  remains controversial. 

 On the other hand, specific effects play important role for the capacitance, especially in the case of polar solvent, where $\epsilon$ may exhibit strong dependence on $\rho$ and on the distance from the electrode
 \cite{Fumagalli2018,Motevaselian2020,Dalessandro2024, Gongadze2015}. In general, the performance of the double layer  capacitance is determined by a combination of  many factors that include  concentration of ions in the electrolyte solution, size and nature of the ions,  solvent polarity, electrode material,   electrolyte--surface interactions,  pore geometry, temperature, etc. \cite{Coskun2024,Kasina2024, Wippermann2023,Zhang2024,COSTA2018214,kornyshev:07:00,fedorov:14:0,deSouza2020,Wu2022,Cats2022,Jeanmairet2022,Gongadze2012,
 Gongadze2015,Cruz2019}.

 It is not easy to disentangle different effects on the capacitance. In this work, we consider the simplest model of ionic systems with the size of the ions taken into account, namely  the restricted primitive model (RPM) of charged hard spheres with equal diameter $a$ and opposite charges in a structureless solvent characterized by the dielectric constant $\epsilon$. In this model, the Coulomb and steric interactions lead to oscillatory decay of the charge density with the distance from a charged boundary~\cite{fedorov:08:1,ciach:18:1,otero:18:0}. We can thus determine the effect of the charge layering near the electrode on the capacitance that should be common for many concentrated ionic solutions. In particular cases, specific interactions and the dielectric constant dependence on local environment  can play an important role. Our purpose, however, is to find a  reference point for the capacitance in concentrated ionic solutions at very small voltages, in analogy with the Debye capacitance for dilute electrolytes.  To achieve this goal, we use the  theoretical approach developed in our previous works for the description of  
 systems with spontaneous inhomogeneity in the bulk, near a charged wall and in a slit geometry \cite{ciach:12:0,ciach:18:1,ciach:21:0,patsahan:22:0,Ciach2023}.

In sec.\ref{sec:theory} we briefly summarize the mesoscopic theory  developed in Ref.\cite{ciach:12:0,ciach:18:1,ciach:21:0,patsahan:22:0,Ciach2023}. In sec.\ref{sec:results_dilute} we present results for dilute electrolytes to test the mesoscopic theory predictions for the capacitance. In sec.\ref{sec:results_dense} we derive our results for the capacitance of systems with large density of ions. We discuss our results in sec.\ref{sec:discussion}, and conclude in sec.\ref{sec:conclusion}.

\section{The mesoscopic theory}
\label{sec:theory}
In this section we briefly summarize the key concepts, assumptions, approximations and results of the mesoscopic theory that can be applied to  dilute as well as to concentrated electrolytes. The theory developed and described in detail  in \cite{ciach:12:0,ciach:21:0,patsahan:22:0,Ciach2023} allows to determine the differential capacitance of the double layer in terms of the structure of the ionic solution. In principle, different levels of approximation in this theory are possible, but in this work we limit ourselves to the simplest approximation to highlight the underlying effect of the structure i.e. the distribution of the ions, on the capacitance. 

In order to calculate the capacitance of the double layer, 
we consider the electrolyte in contact with a planar metallic electrode, and assume that the charge of the electrode is distributed over its flat surface.  In the case of
concentrated ionic solutions, for example water in salt electrolyte, IL, or IL mixture with a neutral solvent, the assumption of point charges is not valid. The simplest model that takes the size of the ions into account is the restricted primitive model (RPM) of charged hard spheres with equal diameter $a$ and opposite charges in a structureless solvent characterized by the dielectric constant $\epsilon$. If the sizes of the  positive and negative charge ions, $a_+$ and $a_-$, are somewhat different, we assume $a=(a_++a_-)/2$.
We adopt this model, and assume in addition that the charge is homogeneously distributed over the volume of the ion. To fix attention, we assume monovalent ions.

 The cases of significant size disparity, different valency of  the cations and the anions  as well as strong specific interactions require separate study. We should mention that significantly different sizes of ionic cores together with specific interactions can lead to spontaneous formation of relatively large charged domains~\cite{borodin:17:0,patsahan:22:1}, as well as to an asymmetric shape of electric double layer capacitance~\cite{fedorov:08:0,Velikonja2015,Igli2019}.

Our theory~\cite{ciach:2023:2} is based on local mesoscopic volume fraction $\zeta_i$ of the $i$-th component of the mixture. The mesoscopic volume fraction is defined by analogy with its  macroscopic counterpart, namely,  $\zeta_i({\bf r})$ is equal to the fraction of a mesoscopic volume with the center at ${\bf r}$ that is occupied by the particles of the considered species. In general, it depends on the scale of the coarsegraining.

\begin{figure}
\includegraphics[scale=0.3]{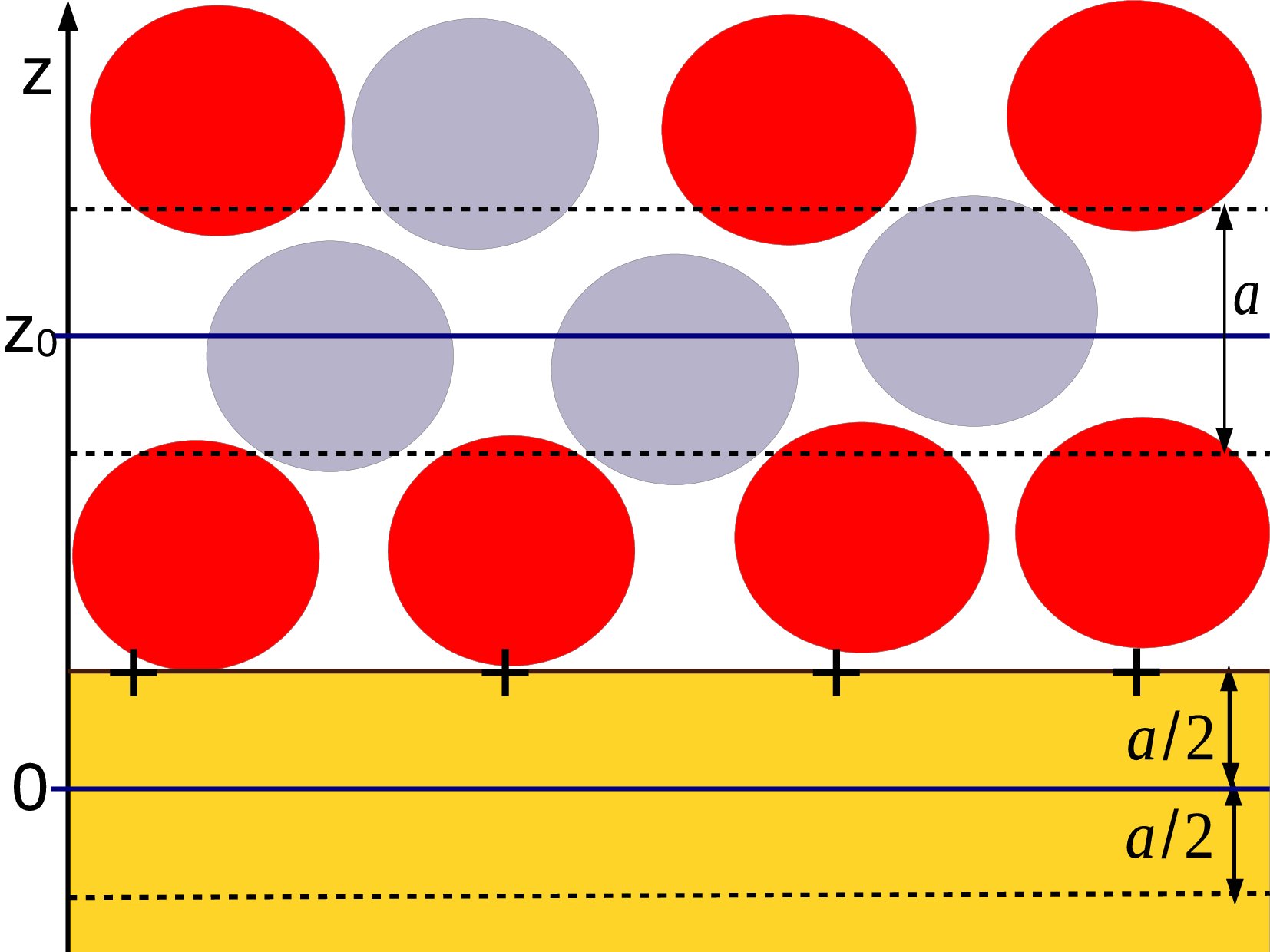}
\caption{Cartoon explaining the mesoscopic densities in our theory in the case of IL. The yellow shaded region represents the solid wall with fixed charge or fixed voltage. The red and gray circles represent the negatively and positively charged ions. The mesoscopic density  is $\rho_i(z)=6\zeta_i(z)/\pi$, where $\zeta_i(z)$ is the fraction of the volume of the layer centered at $z$  and of thickness $a$ that is covered by the $i$-th type ions. A typical layer with the center at $z=z_0$   
is bounded by the dashed lines. With this construction, $\rho_i(z)$ are continuous functions. The charge at the surface of the solid wall is included in the considered system. We  choose $z=0$  at the distance $a/2$ from the surface of the electrode inside the solid, as shown in the cartoon. With this choice, the mesoscopic charge density $c(0)=\rho_+(0)-\rho_-(0)$ contains only the  charge from the wall, i.e. the charge included in the liquid does not contribute to $c(0)$. When the charge of the metallic electrode is confined to the surface at $z=a/2$, a layer of the thickness $a$ with the center at $z<0$ contains no charge, therefore $c(z<0)=0$.  With this formulation of the mesoscopic theory, we assume that the charge in the electrode of area $A$ is equal to $ec(0) A$, and the overall charge neutrality condition is $\int_0^{\infty}dz c(z)=0$. }
\label{fig:meso}
\end{figure}

 We assume that near a flat electrode, $\zeta_i$ depends only on the distance $z$ from the solid-liquid interface.  
As illustrated in Fig.\ref{fig:meso} for the IL in contact with a flat solid surface, we identify the mesoscopic regions with layers of a thickness $a$ that are parallel to the solid surface. The mesoscopic volume fraction of the anions or the cations at the  center of the layer is equal to the fraction of the  volume of that layer that is occupied by the anions or the cations, respectively. With this definition, we obtain continuous functions of the distance from the electrode. Dimensionless mesoscopic densities are defined by $\rho_i(z)=6\zeta_i(z)/\pi$, and in the ionic system, $i=+,-$. It is convenient to introduce the local dimensionless number density of ions, $\rho(z)=\rho_+(z)+\rho_-(z)$ and the local dimensionless charge, $c(z)=\rho_+(z)-\rho_-(z)$. For monovalent ions, the local charge density is $ec(z)$, where $e$ is the elementary charge. 

 There is some ambiguity in defining the mathematical surface representing the solid-liquid interface on the microscopic and mesoscopic level.  We choose the origin of the coordinate frame, $z=0$, inside the solid at the distance $a/2$ from the solid surface at which the electrode charge with the surface charge density $e\sigma_0$ is homogeneously distributed (see Fig.\ref{fig:meso}). From Fig.~\ref{fig:meso} it can be clearly seen that
in our mesoscopic approach,  $c(0)$  
is equal to the dimensionless
surface charge density $\sigma_0^*=a^2\sigma_0$. 
For $z<0$ we have $c(z)=0$ when the charge of the electrode is confined to the surface of the solid, since this surface is now outside the layer with the center at $z<0$. For $0<z<a$, both the electrode and the liquid contribute to $c(z)$. Finally, for $z>a$ the only contribution to the charge density $ec(z)$ comes from the electrolyte. 
The charge-neutrality condition of the whole system in this theory takes the simple form
\begin{equation}
 \int_0^{\infty}dz c(z)=0.
 \label{chne}
\end{equation}
The electrostatic potential at $z=0$ is given by the Poisson equation in the integral form,  
\begin{equation}
\label{Psi0}
\Psi(0)=-\frac{4\pi e}{\epsilon a}\int_0^{\infty} dzzc(z),
\end{equation}
and can be calculated once the shape of $c(z)$ is known. Here and below the distance is in $a$ units. All the characteristic lengths will be in $a$ units as well, i.e. we will consider dimensionless quantities.

Determination of the equilibrium shape of $c(z)$ is the main difficulty of the theory. From thermodynamics we know that in a system with fixed volume and fixed temperature $T$ that is in contact with a bulk reservoir of ions, the grand thermodynamic potential $\Omega$ takes a minimum. Thus, we should consider the grand potential for different forms of $\rho(z)$ and $c(z)$, and find the functions that minimize the functional 
\begin{equation}
\Omega[c,\rho]= U[c]- TS[c,\rho]-\mu N,
\end{equation}
where $\mu$ and $N$ are the chemical potential and the number of the ions, respectively. The internal energy $U[c]$ depends only on the charge $c$ if the specific interactions are neglected, and takes  in $k_BT=1/\beta$ units the form
\begin{equation}
\beta U[c]=\frac{l_B}{2}\int d{\bf r}_1\int d{\bf r}_2 c({\bf r}_1) \frac{g(r)}{r} c({\bf r}_2)
\end{equation}
where $l_B=\beta e^2/a\epsilon$ is the Bjerrum length in $a$ units, $r=|{\bf r}_1-{\bf r}_2|$, and $g(r)$ is the pair distribution function. The entropy $S[c,\rho]$ consists of the entropy of mixing of the anions and the cations, and of a contribution associated with packing of the hard spheres representing the ionic cores.

The exact expressions for $\Omega[c,\rho]$ (including the precise form of $g(r)$ and $S[c,\rho]$) are not known, and different approximate theories were developed. In the bulk, the position independent $\rho$ is a function of the chemical potential and temperature. 
The average charge density is $c=0$ because it is equally probable to find an anion or a cation in a given microscopic volume in the absence of external fields. If an ion is kept at a given position, however, then it is more probable to find oppositely charged ion in its vicinity  than an ion of the same charge, because the energy in the former case is lower. The charge-charge correlation function is a  result of the competition between the energy favoring oppositely charged close neighbors and the entropy favoring random distribution of the ions in the whole volume.

In our mesoscopic theory~\cite{Ciach2023,ciach:2023:2}, we assume that the ions cannot overlap, therefore $g(r)=0$ for $r<1$ ($r$ is in $a$ units), and $g(r)\to 1$ for $r\to \infty$, since at very large distances the ions are not correlated.
In the mean-field (MF) approximation, the correlations are neglected for $r>1$, and  $g(r)\approx\theta(r-1)$, where $\theta(r-1)=1$ for $r>1$ and $\theta(r-1)=0$ for  $r<1$. In this approximation, 
\begin{equation}
\beta U[c]\approx\frac{l_B}{2}\int d{\bf r}_1\int d{\bf r}_2 c({\bf r}_1) \frac{\theta(r-1)}{r} c({\bf r}_2)=\frac{l_B}{2}\int d{\bf k}\hat c({\bf k})\hat V_C(k)\hat c(-{\bf k}),
\end{equation}
where 
\begin{equation}
\label{VC}
\hat V_C(k)=\frac{4\pi \cos(k)}{k^2}
\end{equation}
 is the energy per unit amplitude of a charge wave with the  wavelength $2\pi/k$ excited in the homogeneous system, and $\hat c({\bf k})$ is the amplitude of this wave. $\hat V_C(k)<0$ and the energy decreases when the charge wave with $k>\pi/2$ is excited. It takes a minimum for $k_0\approx 2.46$ (in $1/a$ units), consistent with energetically favorable oppositely charged close neighbors. 

 The charge-charge correlation function in Fourier representation takes in this theory the form 
\begin{equation}
\langle\hat c({\bf k})\hat c(-{\bf k})\rangle = \Bigg[
l_B\hat V_C(k)+\frac{1}{\rho_R}
\Bigg]^{-1}
\label{hVC}
\end{equation}
where $\rho_R=\rho$ in the MF approximation and for $\rho\ll 1$.
 In concentrated solutions, however, $\rho_R<\rho$.
The renormalized density of ions in the mesoscopic theory follows from the energetically favorable charge waves that play a similar role as the  neutral clusters observed in 
concentrated electrolytes~\cite{Goodwin2017,Chen2018,safran:23:0}, i.e. they lead to a smaller density of free ions.
 Equations for $\rho_R$ are developed and discussed in Ref.~\cite{Ciach2023,ciach:2023:2}, 

The charge-charge correlations in real space are obtained by the inverse Fourier transform of  $\langle\hat c({\bf k})\hat c(-{\bf k})\rangle$, and the decay lengths are obtained from simple poles of $\langle\hat c({\bf k})\hat c(-{\bf k})\rangle$ extended to the complex $q$ plane. In general, the imaginary pole $q=i a_i$  gives a monotonic decay with the decay length $1/a_i$, and the complex poles $q=i\alpha_0\pm\alpha_1$ give an oscillatory decay with the decay length $1/\alpha_0$ and the period $2\pi/\alpha_1$.

 In our mesoscopic theory, there are two inverse decay lengths, $a_1>a_2$ for large $T$ and small $\rho$, and the asymptotic decay of correlations at large distances is monotonic.   For $\rho\to 0$ we get $1/a_2\to\lambda_D^*$, where $\lambda_D^*=1/\sqrt{4\pi l_B\rho}=\lambda_D/a$ is the dimensionless Debye screening length (recall that $l_B$ and $\rho$ are dimensionless), and $a_1$ is of order of unity (in $1/a$ units).
 $a_1$ and $a_2$
 merge at the so-called Kirkwood line on the $(\rho,T)$ diagram~\cite{ciach:03:1}, and at lower $T$ they transform to a pair of complex inverse decay lengths that lead together to an oscillatory decay of the charge-charge correlations.
 This result agrees with previous theories and simulations~\cite{attard:93:0,leote:94:0,Torrie1980},  and is consistent with dominant roles of the entropy  and energy  at high and low $T$, respectively. 
 
In a presence of the electrode, the excess density over the bulk value, $\Delta \rho(z)=\rho(z)-\rho_b$, and $c(z)\ne 0 $ minimize the functional $\Delta\Omega[c,\Delta\rho]=\Omega[c, \rho_b+\Delta\rho]-\Omega_{bulk}[0,\rho_b]$. Assuming small voltage or small $\sigma_0$ at the electrode, we expect small $\Delta\rho$ and $c$, and Taylor expand  $\Delta\Omega[c,\Delta\rho]$. When the bulk is in thermal equilibrium, the terms linear in  $\Delta\rho$ and $c$ vanish, and 
$\Delta\Omega[c,\Delta\rho]$ contains terms of the second and higher orders in the fields. In the lowest-order approximation, we keep only the terms quadratic in $\Delta\rho$ and $c$,  and following the steps described in \cite{ciach:18:1,patsahan:22:0} obtain the approximation
\begin{equation}
\beta\Delta\Omega[c,\Delta\rho]=\frac{1}{2}\int dz\Bigg[ c(z) \Bigg(l_B\hat V_C\Big(-i\frac{d}{dz}\Big)+\frac{1}{\rho_R}\Bigg)c(z)+ \Delta\rho(z)\Bigg(R_0-R_2\frac{d^2}{dz^2}\Bigg)\Delta\rho(z)\Bigg]+ h.o.t.
\label{DOVC}
\end{equation}
where h.o.t denotes the terms of higher order in  $\Delta\rho$ and $c$. The parameters $R_0,R_2$ come from the entropic contribution to $\Omega$ and from the  correlations between fluctuations of the local charge~\cite{ciach:21:0,patsahan:22:0}. The differential operator $\hat V_C\Big(-i\frac{d}{dz}\Big)$ is defined through the Taylor expansion.

Minimization of the functional (\ref{DOVC}) leads to  ordinary differential  Euler-Lagrange (EL) equations. When the h.o.t. are neglected, the EL equations for   $\Delta\rho$ and $c$ are  linear and decoupled. The linearized EL equation for $c(z)$ is
\begin{equation}
\Bigg(l_B\hat V_C\Big(-i\frac{d}{dz}\Big)+\frac{1}{\rho_R}\Bigg) c(z)=0.
\label{EL}
\end{equation}
A solution of a linear equation is a sum of exponential terms $\propto \exp(\lambda z)$. In our case, $\lambda=-iq$, where complex $q$ is a solution of the equation $l_B\hat V_C(q)+\frac{1}{\rho_R}=0$.
For our purpose here it is important that the linearized EL equation leads to $c(z)$ decaying  at large distances with the same decay lengths as the charge-charge correlation function (compare (\ref{EL}) and (\ref{hVC})). The solution of (\ref{EL}) must satisfy the charge neutrality condition~(\ref{chne}). 

The equations (\ref{EL}), (\ref{chne}),  (\ref{Psi0}) and $c(0)=a^2\sigma_0$ allow for calculation of the differential capacitance in the case of small voltage, and with the specific effects of the solvent neglected.

\section{results}
\label{sec:results}
\subsection{Capacitance of dilute electrolytes}
\label{sec:results_dilute}
\begin{figure}
\includegraphics[scale=0.45]{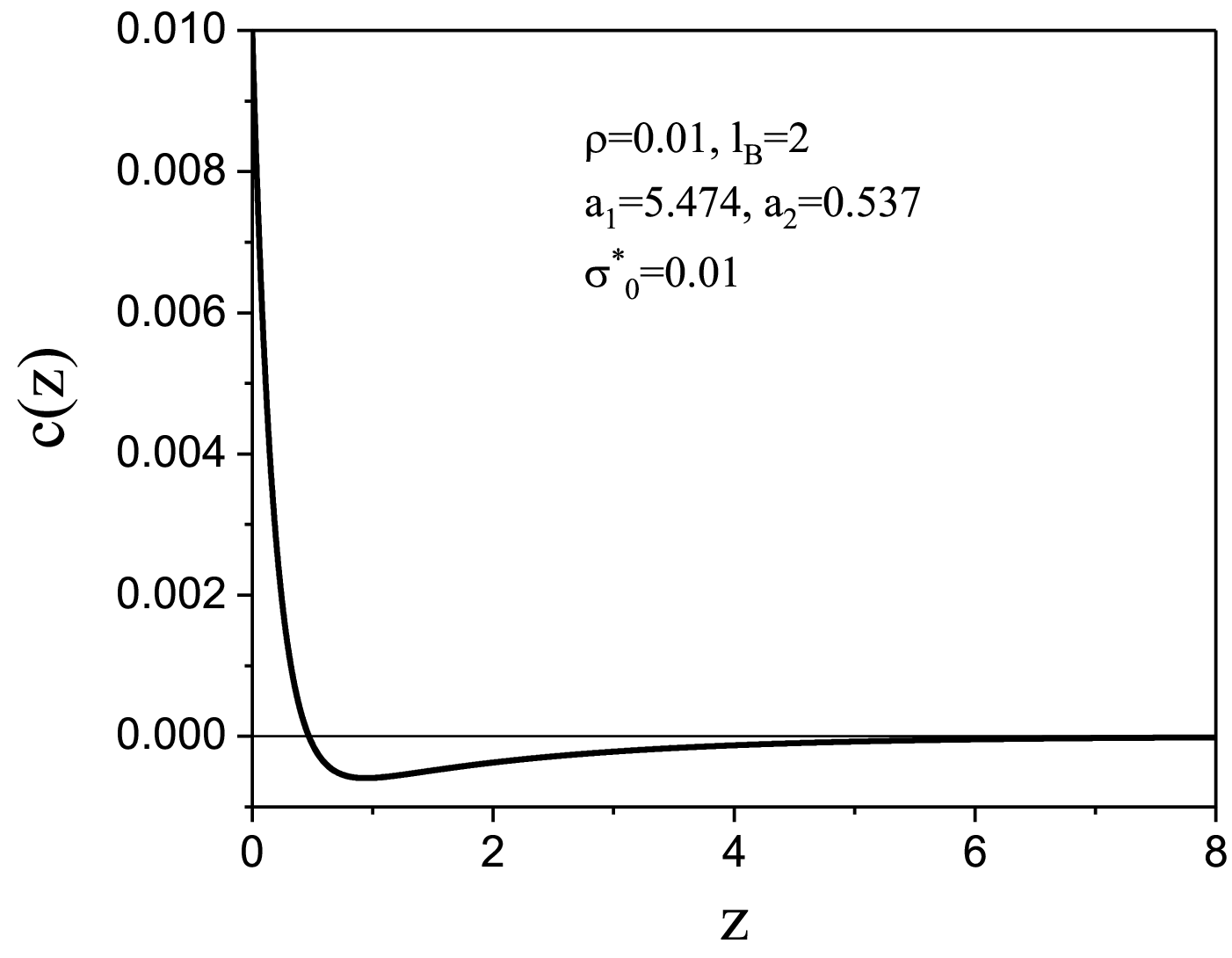}
\caption{The dimensionless charge density in the case of dilute electrolyte with $\rho=0.01, l_B=2$ and dimensionless surface charge density $\sigma_0^*=0.01$ as obtained in the mesoscopic theory. The decay lengths satisfy the equation $4\pi l_B\rho_R \cosh a_i= a_i^2$, and $\rho_R\le\rho$ satisfies the equations presented in ref.\cite{Ciach2023}.  $z$ is in units of the ion diameter $a$.}
\label{fig:cprofile}
\end{figure}

On the high-$T$  small $\rho$ side of the Kirkwood line, the charge-density profile satisfying the charge-neutrality condition  has the form 
\begin{equation}
\label{c(z)}
c(z)=A_1\Big( e^{-a_1z}-\frac{a_2}{a_1} e^{-a_2 z}\Big), 
\end{equation}
where the length unit is the ion diameter $a$. The dimensionless inverse decay lengths are determined numerically based on Ref.\cite{ciach:03:1,Ciach2023}. Similar values were obtained in different theories~\cite{leote:94:0}. A representative $c(z)$ is shown in Fig.\ref{fig:cprofile} for $\sigma_0^*=0.01$, $l_B=2$ and $\rho=0.01$.

The surface-charge density in the mesoscopic theory is 
\begin{equation}
\label{esigma}
e\sigma_0=\frac{ec(0)}{a^2}=\frac{eA_1 (a_2^{-1}-a_1^{-1})a_2}{a^2}.
\end{equation} 
For the potential at $z=0$ we obtain from (\ref{c(z)}) and (\ref{Psi0}) 
\begin{equation}
\label{U}
U=\Psi(0)=\frac{4\pi e A_1(a_2^{-1}-a_1^{-1})}{\epsilon a_1 a}.
\end{equation}
The capacitance $C=d(e\sigma_0)/dU$ is easily obtained from (\ref{esigma}) and (\ref{U}) and is given by
\begin{equation}
\label{Cm}
C_{dil}=\frac{\epsilon a_1a_2}{4\pi a}.
\end{equation}

For dilute electrolytes, $a_2/a\to 1/\lambda_D$ that can be easily seen from the equation $l_B\hat V_C(ia_2)+1/\rho=0$ for the imaginary simple pole  and 
 $\hat V_C(ia_2)\to -4\pi /a_2^2 $  for $\rho\ll 1$ (see (\ref{hVC}) and (\ref{VC})), giving  $a_2^2=4\pi l_B\rho $. Thus, for dilute electrolytes we obtain 
\begin{equation}
\label{Cma}
C_{dil}=\frac{\epsilon a_1}{4\pi \lambda_D},
\end{equation}
where the dimensionless parameter $a_1$ in our theory is larger than 2 and increases with increasing $(l_B\rho)^{-1}$, but remains of order of unity. Precise value of this inverse microscopic length should be determined in a more exact microscopic theory. Thus,  in the limit of  dilute electrolytes we obtain the Debye capacitance  up to a parameter $a_1=O(1)$. To compare our predictions with the results of simulations or experiments on the quantitative level, however, we should take  into account that the formulas (\ref{Cm}) and (\ref{Cma}) contain the  factor $a_1$ that in the theory with the microscopic structure averaged over the region with the linear size $a$ is $2\le a_1\le 6$ rather than $a_1=1$ present in the Debye capacitance.

\subsection{Capacitance of concentrated electrolytes}
\label{sec:results_dense}
\begin{figure}
\includegraphics[scale=0.33]{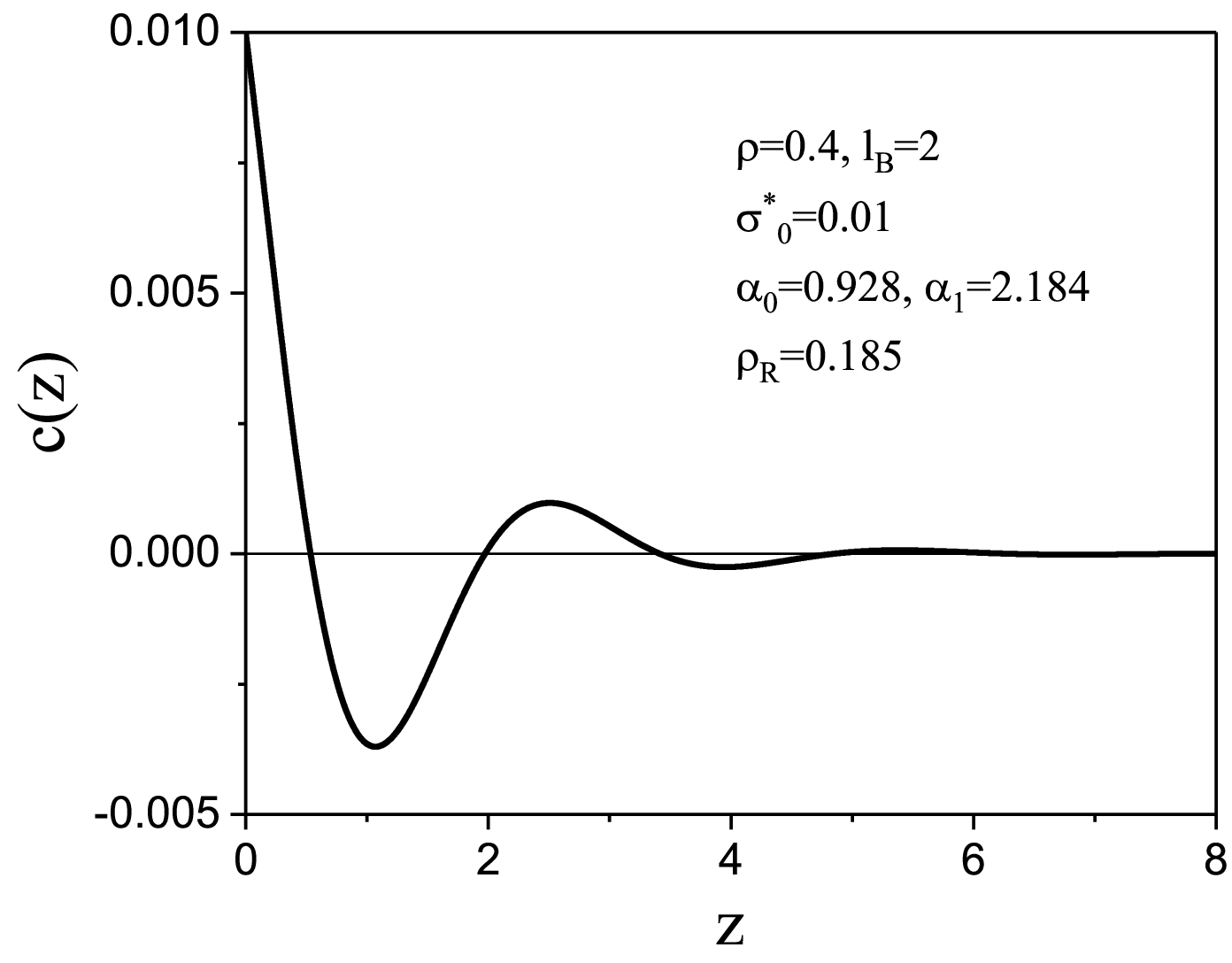}\includegraphics[scale=0.33]{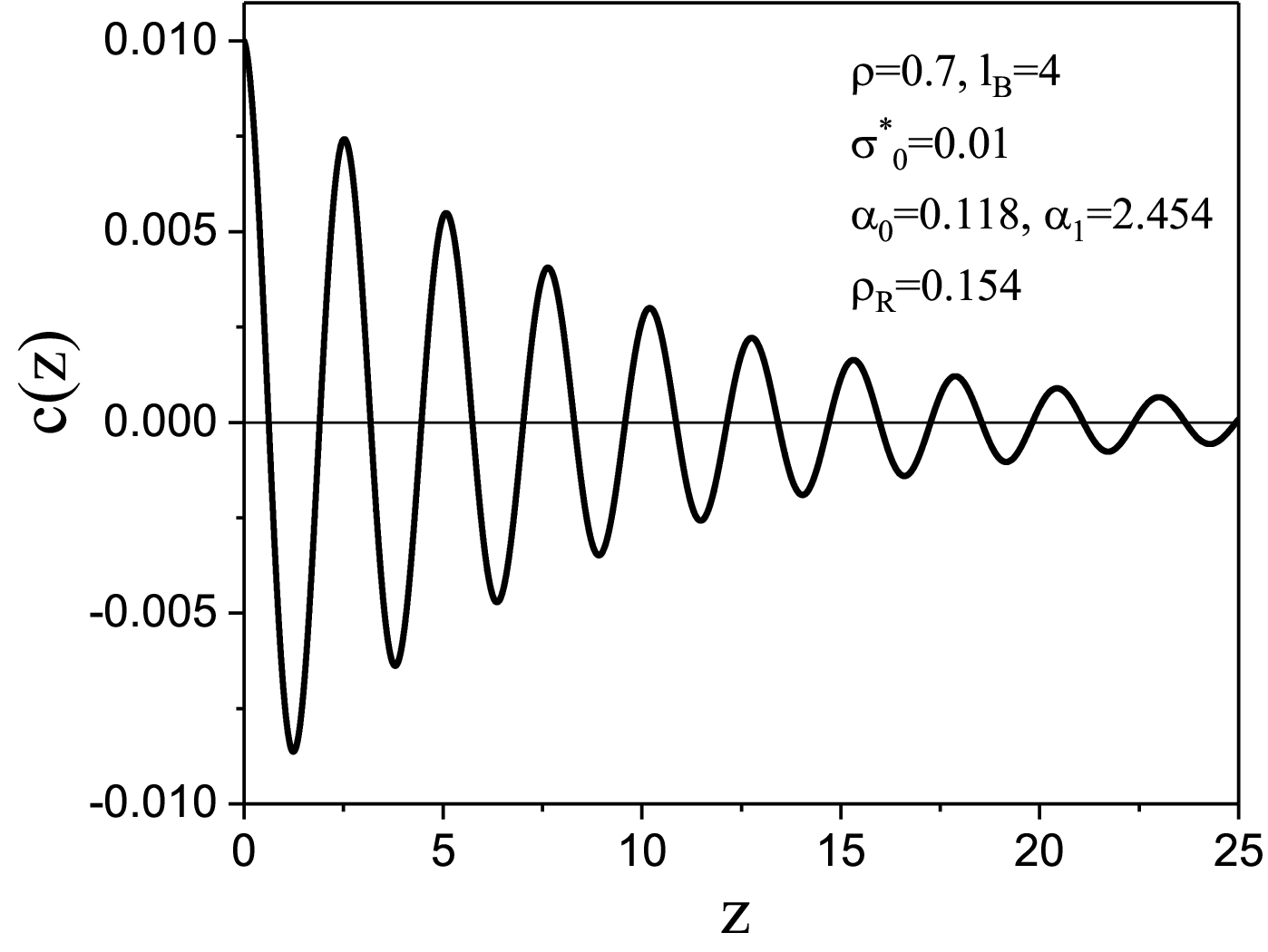}
\caption{The dimensionless charge density in the case of the concentrated electrolyte with $\rho=0.4, l_B=2$ and  $\rho_R\approx 0.185$ (left) and  $\rho=0.7, l_B=4$ and  $\rho_R\approx 0.154$  (right) for dimensionless surface charge density $\sigma_0^*=0.01$, as obtained in the mesoscopic theory. $\alpha_0$ and $\alpha_1$  satisfy the equations $(\alpha_1^2-\alpha_0^2)+4\pi l_B\rho_R \cosh \alpha_0\cos \alpha_1=0$ and $\alpha_1\alpha_0-2\pi l_B\rho_R\sinh\alpha_0\sin\alpha_1=0$~\cite{Ciach2023}.  $z$ is in units of the ion diameter $a$.}
\label{fig:cprofileo}
\end{figure}

The solution of the EL equation (\ref{EL}) at the low-$T$ side of the Kirkwood line has the form
 \begin{equation}
 \label{c0}
 c(z)=A_ce^{-\alpha_0z}\sin(\alpha_1z+\theta).
 \end{equation}
 The above dimensionless charge density profile agrees with the fixed surface charge density $\sigma_0$ and satisfies the charge-neutrality condition (\ref{chne}) when 
 \begin{equation}
\label{calA}
A_c=\frac{a^2\sigma_0}{\sin \theta},
\hskip1cm  \theta=\arctan\Big(-
\frac{\alpha_1}{\alpha_0}
\Big).
\end{equation}
A typical dimensionless charge density near a weakly charged electrode is shown in Fig.\ref{fig:cprofileo}. Predictions of the mesocopic theory for the charge profile~\cite{ciach:18:1} agree  with simulation results of the RPM model in the case of large density of ions~\cite{fedorov:08:1}. In addition, as shown for example in Ref.\cite{otero:18:0},  Eq.(\ref{c0}) perfectly fits the charge density in atomistic simulations of IL - alcohol mixture for  $z>2\pi/\alpha_1$. The atomistic simulations give different shapes of $c(z)$ for $z<2\pi/\alpha_1$, 
because the  microscopic charge density obtained in the simulations  is averaged over the layers of the thickness $a$ in the mesoscopic theory. 

The electrostatic potential (\ref{Psi0}) at $z=0$ for the charge density given by (\ref{c0}) and (\ref{calA}) is 
\begin{equation}
\label{Psiexp}
U=\Psi(0)=-\frac{4\pi e A_c\alpha_1\cos(\theta)}{\epsilon a\alpha_0(\alpha_0^2+\alpha_1^2)}.
\end{equation}
The capacitance can be easily calculated using $e\sigma_0= eA_c\sin\theta/a^2$ with Eqs. (\ref{calA}) and (\ref{Psiexp}), and the result is
\begin{equation}
\label{C}
C=\frac{\epsilon  (\alpha_0^2+\alpha_1^2)}{4\pi a}=\frac{\varepsilon_0\varepsilon_r  (\alpha_0^2+\alpha_1^2)}{ a},
\end{equation} 
where  $\epsilon= 4\pi\varepsilon_r\varepsilon_0$, with $\varepsilon_r$ and $\varepsilon_0\approx 9\cdot 10^{-6}\mu F/m$ denoting the relative dielectric constant and the vacuum permittivity, respectively.

The dimensionless wavenumber $\alpha_1\le k_0\approx 2.46$ of the damped charge oscillations  decreases slightly for increasing $(l_B\rho)^{-1}$ in the concentrated electrolyte~\cite{ciach:03:1}. As found recently~\cite{smith:16:0,lee2017,ciach:2023:2}, in concentrated electrolytes the dimensionless  inverse decay length is $\alpha_0\ll 1$, hence we can assume that in concentrated electrolytes and IL $\alpha_0^2\ll\alpha_1^2$, and
\begin{equation}
\label{Ca}
C\approx\frac{\varepsilon_0\varepsilon_r \alpha_1^2}{ a}.
\end{equation}  
The formulas (\ref{C}) and (\ref{Ca}) relating the capacitance with the period $2\pi a/\alpha_1$ of the damped charge oscillations near the electrode are the main result of this work.

\section{discussion}
\label{sec:discussion}
\begin{figure}
\includegraphics[scale=0.4]{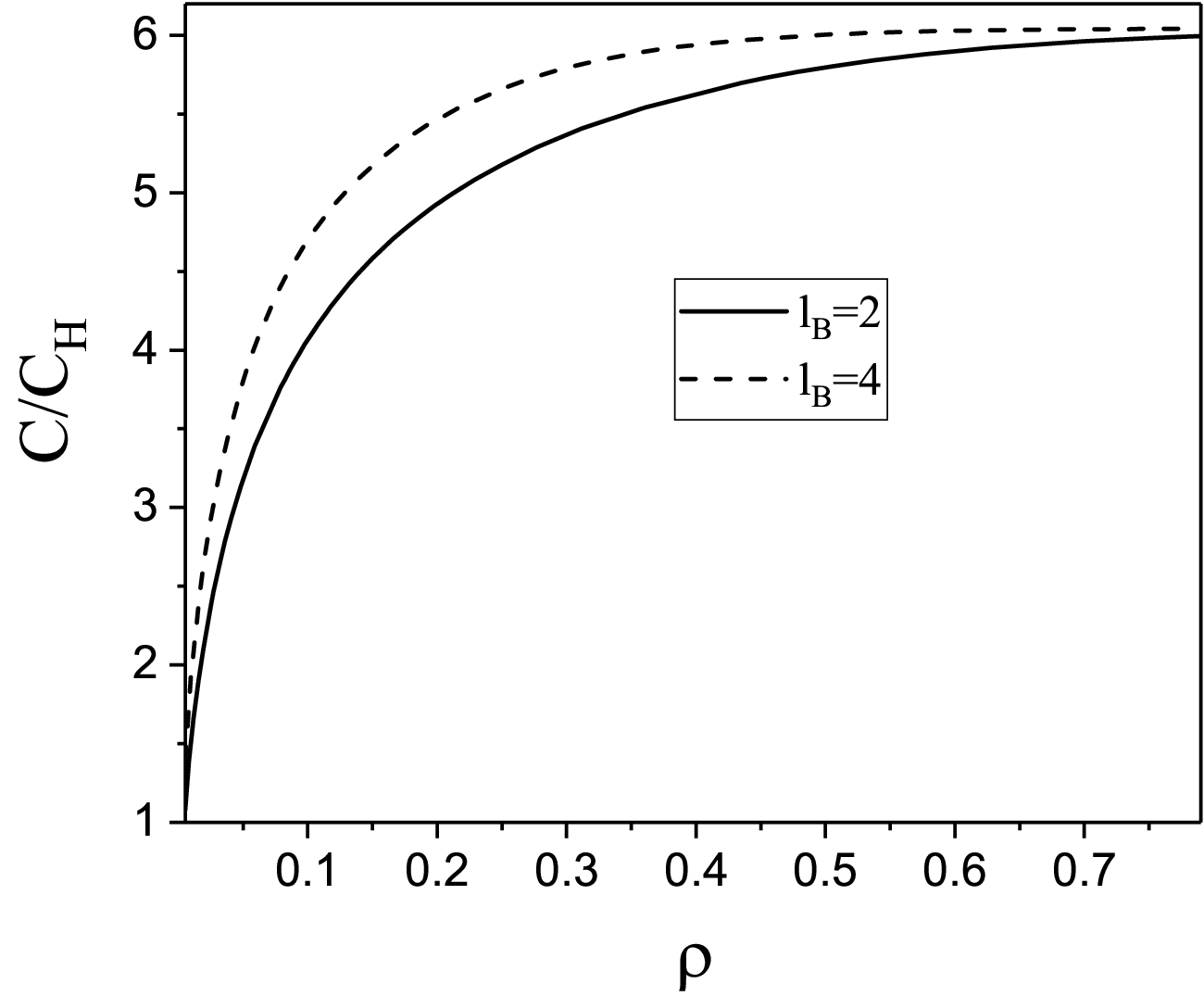}
\caption{The capacitance $C/C_H$ obtained from (\ref{Cm}) and (\ref{C}) as a function of the dimensionless density of ions for the Bjerrum length $l_B=2$ and $l_B=4$ in $a$-units. }
\label{fig:Cr}
\end{figure}

The expressions for the capacitance of dilute and concentrated electrolytes in the RPM near a flat metallic electrode, Eqs.~(\ref{Cm}) and ~(\ref{C}), are significantly different. They  become identical, however, at the Kirkwood line separating the monotonic and oscillatory asymptotic decays of the charge density, because at the Kirkwood line $a_1=a_2=\alpha_0$ and $\alpha_1=0$. Thus, we obtained a continuous function for the whole range of the density of ions. The dependence of $C/C_H$, where $C_H=\varepsilon_r\varepsilon_0/a$, on the density of ions for fixed Bjerrum lengths $l_B=2$ and $l_B=4$  is shown in Fig~\ref{fig:Cr}.  

Our general analytical formulas (\ref{Cm}) and ~(\ref{C}) were obtained from the linearized EL equation (\ref{EL}), therefore they are valid only for very small voltage. Nevertheless, they highlight the effect of the charge distribution on the capacitance for the whole range of the density of ions on a general qualitative level.
For large voltages, the charge density is no longer small, and the coupled nonlinear EL equations for $c$ and $\Delta\rho$ (see Ref.\cite{ciach:2023:2}) have to be solved to determine $C$. This is possible only numerically for particular cases, and will be a subject of our future study. 

The formulas for the capacitance become particularly simple for very dilute and very dense electrolyte. In the former case, we obtain the well known Debye capacitance $C_D=\epsilon/(4\pi\lambda_D)$ up to a dimensionless coefficient of order unity (see Eq.(\ref{Cma})). For the very concentrated electrolyte or IL, we obtain the formula (\ref{Ca}) that is strikingly similar to the Helmholtz capacitance $C_H(L)=\epsilon/(4\pi L)$ in the early model of the double layer, where $L$ is the distance between the electrode and the surface occupied by the counterions. 
 We take into account the whole oscillatory  charge profile such as the ones shown in Fig.\ref{fig:cprofileo}, and find that the alternating oppositely charged layers have the same effect as a single layer of counterions located at the distance  $L=a/\alpha_1^2$ from the electrode. This shows that the simplest model of the double layer works well even in the case of a rather complex structure, but with the distance of  the virtual monolayer of counterions from the electrode, $L$, depending on the dimensionless wavenumber $\alpha_1\approx 2.46$ of the charge oscillations.

Let us compare predictions of 
the mesoscopic theory for the charge-density profiles shown in Fig.\ref{fig:cprofileo} with the classical Debye capacitance. 
For $\rho=0.4$ and $l_B=2$, and the diameter of hydrated ions $ a\approx 0.5 nm$ corresponding to 
$\sim 2.65$M $NaCl_{aq}$, we get $\alpha_0\approx 0.93$  and $\alpha_1\approx 2.18 $, and obtain from (\ref{C})  
\begin{equation}
\label{C1}
C\approx 10\varepsilon_r \mu F/cm^2. 
\end{equation}
From $\lambda_D=a/\sqrt{4\pi l_B\rho}$ we get  in this case $\lambda_D\approx 0.16 nm$, and the formula valid for dilute electrolytes gives
\begin{equation}
\label{Cdil1}
C_{D}\approx 5.6\varepsilon_r \mu F/cm^2. 
\end{equation}
The Debye length  in this case differs from the physically relevant lengths $a/\alpha_0\approx 0.54 nm$ and $2\pi a/\alpha_1\approx 1.44 nm$, however.  
In another example shown in Fig.\ref{fig:cprofileo} with $\rho=0.7$ and $l_B=4$,  we obtain assuming $a=0.9 nm$
\begin{equation}
\label{Cdil2}
C\approx 6\varepsilon_r \mu F/cm^2. 
\end{equation}
The Debye length is $\lambda_D\approx 0.15nm$, and 
\begin{equation}
\label{Cdil3}
C_{D}\approx 5.9\varepsilon_r \mu F/cm^2. 
\end{equation}
The values of $C$ and $C_D$ are very similar in this case, but the interpretation is quite different. For the oscillatory decay of the charge density, the key factor is the period of the charge oscillations near the electrode, and the Debye length is not associated with characteristic  lengths of the charge distribution. These examples show that care must be taken in interpreting experiments and simulations, because correct numbers can follow from incorrect formulas.

In order to verify the accuracy of  $C$ given in (\ref{C}) and (\ref{Ca}) on the quantitative level,  we should compare  the theoretical and simulation results for the same model.
In Ref.~\cite{fedorov:08:1} $C$ was obtained by simulations of the RPM with $a= 1nm, T=450 K, \epsilon_r=2$ and $\rho\approx 0.6 nm^{-3}$, and the result of simulations was $C/C_D\approx 0.15$. For the above parameters we obtain 
$\lambda_D\approx  0.084nm$ and  (\ref{Ca}) gives
$
C/C_D=\alpha_1^2\lambda_D/a\approx 0.5,
$
where we used $\alpha_1=2.45$. Recall that
 the capacitance obtained in the mesoscopic theory for dilute electrolytes was overestimated by the factor $2\le a_1\le 6$. In the case of $\rho=0.6nm^{-3}$ and large $l_B$, the theoretical result is about three times larger than the simulation result, i.e. a systematic overestimation of the capacitance  by a factor $\sim 3$ 
 is present in the mesoscopic theory for the whole density range.
 
  Our results  show  the effect on the capacitance of the charge ordering in the region extending to large distances from the electrode. On the quantitative level, however, the capacitance depends also on  the details of the microscopic structure in the vicinity of the electrode that should be determined within a  more exact microscopic theory. In our mesoscopic theory the effect of the microscopic structure can be taken into account by additional scaling factor that based on the comparison with simulations is about $1/3$. 


Let us finally discuss consequences of (\ref{Ca}) on a general level. The capacitance decreases with increasing size of the ions, in agreement with experimental results for aqueous ionic solutions and IL~\cite{Lockett2010,Dalessandro2024}. The dimensionless period of the charge wave in concentrated electrolytes depends on density rather weakly.
According to our mesoscopic theory,  in IL or highly concentrated electrolytes the dimensionless $\alpha_1=2\pi a/\lambda_c$ is $2\le \alpha_1\le 2.46$, corresponding to the wavelength of the  charge-density wave $2.55 a\le \lambda_c\le 3.14a$. Hence, in IL we have for (\ref{Ca}) the approximation
\begin{equation}
\label{Ca1}
\frac{4\varepsilon_r\varepsilon_0}{ a}\le C\le\frac{6\varepsilon_r\varepsilon_0}{ a}.
\end{equation}
In the RPM, the dependence of $\varepsilon_r$ (and in turn of $l_B$) on the density of ions and on the distance from the electrode is neglected, whereas in different solvents, especially in water, this dependence can be quite strong. 
 In aqueous solutions $ \varepsilon_r$ decreases from about $80$ in pure water to about $40$ for $5M$ solution of $NaCl$~\cite{smith:16:0}. In the Stern layer, the orientations of dipoles of water molecules in the hydration shells of ions are almost fixed, and the dielectric constant may decrease  to $\varepsilon_r\sim  5$ or even less~\cite{Velikonja2014,Dubtsov2018,Igli2019,Park2024}. 
 Thus,  quantitative predictions for the capacitance in particular cases are not possible within the RPM, especially for polar solvents such as water. 
 Our results show, however, the general relation between the capacitance and the period of the damped charge oscillations. The complex charge distribution can be replaced by the simplest model of the double layer, provided that the virtual single layer of counterions is separated from the electrode by the distance equal to the diameter of the ions re-scaled by the coefficient proportional to  $\alpha_1^{-2}$, where $\alpha_1$ is the wavenumber of the damped charge oscillations  in $1/a$ units and the proportionality constant is $\sim 3$. 

\section{conclusion}
\label{sec:conclusion}
Our goal was to determine on a very general level the effect of charge ordering in concentrated electrolytes and IL on the capacitance  of the double layer.
We limited ourselves to the restricted primitive model,  RPM, where spherical ions with equal diameters and opposite charges are dissolved in structureless solvent characterized by the dielectric constant $\epsilon$. This way we can determine the effect of the Coulomb and steric interactions in the absence of  specific effects that differ from one system to the other.
We  
obtained very simple expressions for the capacitance in dilute and concentrated electrolytes in the framework of the same mesoscopic theory.
The main conclusion is that the simplest early model of the double layer works surprisingly well in the case of large density of ions, provided that the distance between the virtual single layer of counterions and the electrode is equal to the ion diameter re-scaled by a coefficient determined by the period of the damped charge oscillations. This conclusion agrees with recent simulation results~\cite{Park2024}.   
Our formulas (\ref{C}) and (\ref{Ca}) should play for concentrated ionic systems a similar role as the Debye capacitance plays for dilute electrolytes,  i.e. they can serve as a reference point  that allows to disentangle universal and specific effects on the capacitance.

\section*{Acknowledgments}
We gratefully acknowledge the financial support from the European Union Horizon 2020 research 
and innovation programme under the Marie
Sk\l{}odowska-Curie grant agreement No~734276 (CONIN). 

%

\end{document}